\documentclass[preprint,doublecol]{epl2}

\usepackage{epsfig,graphicx,graphics,times}

\usepackage{amssymb}
\usepackage{amsmath}
\usepackage{graphicx}
\usepackage{dcolumn}
\usepackage{bm}

\begin{document}

\title{Geometric-Phase-Effect Tunnel-Splitting Oscillations in Single-Molecule Magnets with Fourth-Order Anisotropy Induced by Orthorhombic
Distortion}
\shorttitle{Geometric-Phase-Effect Tunnel-Splitting Oscillations in Single-Molecule Magnets\ldots}

\author{ M. S. Foss-Feig\thanks{Current address:  Department of Physics, University of Colorado, Boulder, CO   80309} and Jonathan R. Friedman\thanks{E-mail:\email{ jrfriedman@amherst.edu}}}
\shortauthor{M. S. Foss-Feig and Jonathan R. Friedman}
\institute{
Department of Physics, Amherst College, Amherst, MA 01002-5000}

\date{\today}

\abstract{
We analyze the interference between tunneling paths that occurs for a spin system with both fourth-order and second-order transverse anisotropy.
Using
an instanton approach, we find that as the strength of the second-order transverse anisotropy is increased, the tunnel splitting is modulated, with
zeros occurring periodically.  This effect results from the interference of four tunneling paths connecting easy-axis spin orientations and occurs in
the absence of any magnetic field.
}


\maketitle

Geometric-phase effects play an important role in spin dynamics.  Notably, a geometric phase lies at the heart of spin-parity effects, such as Kramers
degeneracy \cite{149, 150, 599}.  Tunneling of a spin (or magnetic particle) between degenerate orientations can be modulated by such geometric-phase
effects via the interference between multiple tunneling paths, with complete suppression (or ``quenching") of the tunnel splitting occurring when
tunneling paths interfere destructively.

Garg has considered the case in which the tunneling of a spin with biaxial anisotropy can be modulated by the application of a magnetic field along the hard
axis \cite{166, 601}.  Specifically, he studied the Hamiltonian
\begin{equation}
\mathcal{H}=D\left(J^2-\hat{J}_z^2 \right)+E\left(\hat{J}_x^2-\hat{J}_y^2\right)-h_x\hat{J}_x,
\label{HamGarg}
\end{equation}
for spin $J$ and anisotropy constants $D>E>0$ and $h_x\equiv g \mu_B H_x$.  He showed periodic quenching of the tunnel splitting as a function of
magnetic field $H_x$ due to geometric-phase interference between tunneling paths.  Several extensions and variants of this problem have been studied
theoretically \cite{601,248,609,246,610,592, kg2, 597, 414, 593, 570, 607, 594,648}.  In a ground-breaking experiment, Wernsdorfer and Sessoli \cite{162} found
clear
evidence of this effect in the single-molecule magnet Fe$_8$, which is reasonably well described by Eq.~\ref{HamGarg}.

The case originally studied by Garg involved no magnetic field along the longitudinal (z) direction.  Tunneling can occur between the metastable
ground
state $|J\rangle$ and an excited state $|-J+n\rangle$ when the applied field brings them into resonance, i.e. when $h_z\equiv g \mu_B H_z = nD, n=0,
1,
2, \ldots, 2J-1$ \cite{33}.  In Fe$_8$, it was discovered that at these resonance longitudinal fields, a transverse field along the hard axis
continued
to produce interference effects \cite{162}.  Thus, at certain values of $h_z$ and $h_x$, destructive interference occurs, suppressing tunneling.
These
so-called diabolical points in $h_z$-$h_x$ parameter space have been studied theoretically by several authors using different approaches \cite{601, 248, 609, 592,kg2 ,
607, 648}.

The Mn$_{12}$ single-molecule magnet has a four-fold transverse magnetic anisotropy and, like Fe$_8$, displays resonant tunneling between two
easy-axis
orientations \cite{33}.  One variant of this molecule, Mn$_{12}$-tBuAc, appears to display the four-fold symmetry with high accuracy \cite{571,600,645}.
Here we consider the interference that occurs between tunneling paths in such a system and how that interference can be modified by the presence of a
second-order transverse anisotropy perturbation.  Such a perturbation could potentially be induced by the application of uniaxial pressure to a sample
of Mn$_{12}$.  We find that the tunnel splitting is periodically quenched as a function of the strength of the perturbation.  This interference effect
takes place \emph{in the absence of any magnetic field.}  In addition, we find a pattern of diabolical points in parameter space when a longitudinal
field $h_z$ is applied.

Several investigations have studied the inclusion of a fourth-order transverse anisotropy as a perturbation in Eq.~\ref{HamGarg} that affects the number and positions of quenches produced by the external field \cite{601, kg2, 570, 607,162}.  Other work has studied how a transverse magnetic field modulates the interference among the four tunneling paths defined by a fourth-order transverse anisotropy \cite{597,414}. In the present study the fourth-order anisotropy term is the primary transverse anisotropy in the problem, producing four interfering paths, but the interference is modulated by the strength of the second-order anisotropy, leading to periodic quenches as that term is varied. In particular, we consider a spin governed by the Hamiltonian
\begin{equation}
\mathcal{H}=D\left(J^2-\hat{J}_z^2 \right)+k\left(\hat{J}_+^4+\hat{J}_-^4\right)+
\lambda\left(\hat{J}_x^2-\hat{J}_y^2\right),
\label{Ham}
\end{equation}
where the $\hat{J}$'s are standard spin operators.  We restrict the values of $k$ and $\lambda$ such that the z axis is the spin's easy axis (see Appendix).

We calculate the ground-state tunnel splitting using an instanton approach, valid in the semi-classical limit of large $J$.  Since we are considering
tunneling between energy minima, there can be no classical trajectories connecting the endpoints of the motion.  The quantum trajectories with the
largest amplitude are the instantons, obtained from the Euclidian (imaginary-time) Lagrangian
\begin{equation}
L_E=L(t\rightarrow -i\tau)=iJ\left(\cos\theta-1\right)\frac{d\phi}{d\tau}-\mathcal{H}(\theta,\phi),
\label{le}
\end{equation}
where $\mathcal{H}(\theta,\phi)$, the Hamiltonian expressed in terms of spherical angles, is identified with the spin's classical energy $E_{cl}(\theta,\phi)$ defined in Eq.~\ref{Ecl} below. Since the instanton paths are extrema of the action in imaginary time, the total energy must be a conserved quantity along each path.

The ground-state (zero-temperature) tunneling rate is given by
\begin{equation}
\Delta=\sum_i\gamma_i e^{-S_{E,i}}\;,
\label{instanton splitting}
\end{equation}
where the sum is over all instantons and the pre-factors $\gamma_i$ arise from integrating the fluctuations around the instanton paths \cite{23}.  The
Euclidean action $S_E$ for an instanton path is
\begin{eqnarray}
S_E&=&-\int_{\tau_i}^{\tau_f} L_E d\tau \nonumber \\
&=&iJ\int_{\phi(-\infty)}^{\phi(+\infty)}(1-\cos\theta)d\phi+\int_{\tau_i}^{\tau_f}\mathcal{H}(\theta,\phi)d\tau
\end{eqnarray}
The first term has the form of a Berry phase:  The difference in the imaginary part of the action between two paths is proportional to the real part of
the solid angle (on the complexified unit sphere) circumscribed by the two paths.  Since the last term is the total energy, it is a constant for an
instanton path and, by construction, zero for the Hamiltonian, Eq.~(\ref{Ham}).

\begin{figure}[htbp]
\centering
\includegraphics[width=0.48\textwidth]{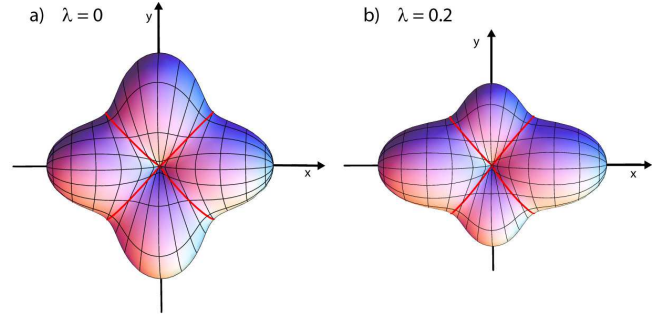}
\caption{(Color online) Three-dimensional plot of energy landscape for a spin governed by Eq.~\ref{Ecl} with $a\equiv k J^2=0.1$ and $\lambda=0$ (a) or $\lambda=0.2$ (b).  The energy minima are along the z axis (into and out of the page) while the x and y axes are the hard axes.  The solid red curves show the four instanton paths with only the real parts of $\theta$ and $\phi$ displayed.  These paths were numerically calculated using Eqs.~\ref{p} and \ref{pdot}.  A rather large value of $a$ is used here to ease visual representation.  For Mn$_{12}$, $a\approx 3 \times 10^{-3}$.}
\label{Effects of Lambda}
\end{figure}

For the case of Eq.~(\ref{Ham}), the spin dynamics can be understood in terms of four instanton paths along the complexified unit sphere. When
$\lambda=0$, the spin's energy has four saddle points located in the x-y plane, with instanton paths connecting the +z and -z directions virtually
``passing through'' these saddle points, as shown  in Fig.~\ref{Effects of Lambda}a, where the solid red curves show numerically calculated instanton paths with only the real parts of $\theta$ and $\phi$ displayed.  For  $\lambda>0$, the energy landscape is
changed, as shown in Fig.~\ref{Effects of Lambda}b, and the instanton paths are pulled towards the y axis.  Reflection symmetry in
the x-z and y-z planes requires that all four paths have the same amplitude, but each has a different geometric phase that leads to interference.  We
therefore concern ourselves with calculating the imaginary part of the action.  The interference effect can be understood by considering, say, the top
two paths  in Fig.~\ref{Effects of Lambda}.  The solid angle between these paths decreases with increasing $\lambda$, modulating the interference
between their tunneling amplitudes.  An identical effect occurs for the bottom two paths.

For ease of computation we measure energy in units of $D$ (i.e.~we set $D=1$).  We choose slightly unconventional spherical coordinates, with the
polar
angle $\theta$ measured from the x axis and the azimuthal angle $\phi$ measured from the y axis to avoid having coordinate singularities at the poles.
In terms of these angles, the classical energy associated with Eq.~(\ref{Ham}) is
\begin{eqnarray}
E_{cl}(\theta,\phi)&=&J^2(1-\sin^2\theta\sin^2\phi)\nonumber \\
& &+\lambda J^2(\cos^2\theta-\sin^2\theta\cos^2\phi)\nonumber \\
& &+2kJ^4(\cos^4\theta+\sin^4\theta\cos^4\phi\nonumber \\
& &-6\cos^2\theta\sin^2\theta\cos^2\phi), \label{Ecl}
\end{eqnarray}
which has two degenerate absolute minima (with $E_{cl}=0$) at $\theta=\pi/2,\phi=\pm\pi/2$.

We can quickly gain a large amount of information about the instantons by solving Eq.~(\ref{Ecl}) for the $\theta(\phi)$ paths that conserve energy.
Making the substitution $p=J\cos\theta$, the conjugate momentum to $\phi$, the instantons must satisfy the constraint $E_{cl}=0$, which has four
solutions:
\begin{equation}
p_{\pm,\pm}(\phi)=\pm J\sqrt{\frac{A(\phi)\pm i\sqrt{B(\phi)}}{C(\phi)}},
\label{p}
\end{equation}
where the subscripts refer to the choices of signs and

\begin{eqnarray}
A(\phi) &=& 4a\cos^4\phi+12a\cos^2\phi-\sin^2\phi-\lambda\cos^2\phi-\lambda \nonumber \\
B(\phi) &=& -\left(128a^2+16a(\lambda-2)+(\lambda-1)^2\right)\cos^4\phi \nonumber \\
  & & +2\left(1 - \lambda^2 + 8a(\lambda+2)\right)\cos^2\phi-(\lambda+1)^2\nonumber \\
C(\phi) &=& 4a(\cos^4\phi+6\cos^2\phi+1),
\end{eqnarray}
with $a\equiv k J^2$.

Two of the solutions in Eq.~(\ref{p}), $p_{\pm,+}$, do not have values of $p=0$ at the endpoints $\phi=\pm\pi/2$.  While such "boundary-jump instantons"
\cite{kg2} are allowed solutions, numerical evaluation of these solutions shows that their action is much larger than for the other two solutions and
therefore they give negligible contribution to the tunnel splitting.  We neglect them in further discussion.

\begin{figure}[htbp]
\centering
\includegraphics[width=0.450\textwidth]{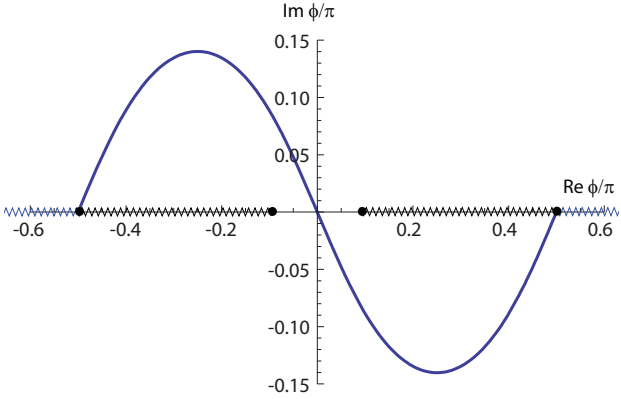}
\caption{Instanton contour in the complex-$\phi$ plane for $a=10^{-4}$ and $\lambda=0$.  Branch points and branch cuts are shown.  The contour changes
only slightly as $a$ is varied by orders of magnitude.}
\label{ParametricG3}
\end{figure}

The other solutions $p_{\pm,-}$ for $0<a<1/2$ and $|\lambda|<\lambda_c \equiv 4\sqrt{a-2a^2}$ have four branch points along the real-$\phi$ axis at
$\phi=\pm\pi/2$ and at the zeros of $B(\phi)$.  Evaluating the action along the real-$\phi$ axis is then potentially problematic and the
integration contour must be chosen with some care.  A judicious choice of branch cuts is shown in Fig.~\ref{ParametricG3}.  We find the instanton
trajectories by considering the (imaginary) time dependence of the coordinates.

The equations of motion for the variables $\phi$ and $p$ can be found using Hamilton's equations:

\begin{eqnarray}
\dot{\phi}&=&-2iJp(4ap^2+J^2\sin^2\phi+ \lambda J^2(1+\cos^2\phi)+\nonumber \\
&&12a(2p^2-J^2)\cos^2\phi+ 4a(J^2-p^2)\cos^4\phi)/J^2\label{phidot}  \\
\dot{p}&=&-i\sin(2\phi)(J^2-p^2)(2a(J^2-7p^2)+J^2(1-\lambda)+\nonumber \\
&&2a(J^2-p^2)\cos(2\phi))/J^2.
\label{pdot}
\end{eqnarray}
Eq.~(\ref{phidot}) can be decoupled by substituting one of the solutions $p_{\pm,-}$ for $p$.  The equation can then be solved numerically to find the real and
imaginary parts of $\phi(\tau)$.  For each choice of $p_{\pm,-}$ we find two solutions, one with endpoints at $\phi=\pm\pi/2$ (passing through
$\phi=0$) and the other with endpoints $\phi=\pi/2$ and $3\pi/2$ (passing through $\phi=\pi$) \footnote{While our numerical calculations show that
each
is a solution, we have not proven that these are the \emph{only} solutions.  However, we do not expect more than four instanton solutions with equal
magnitude actions given the symmetry of the problem.}.  Four more solutions can be found by noting that Eqs.~(\ref{phidot}) and (\ref{pdot}) are invariant
under the transformation $(\phi\rightarrow-\phi^*, p\rightarrow p^*)$ followed by complex conjugation of the equations.  This yields a total of eight
solutions, four of which are instantons and four anti-instantons.  These can be sorted by using Eq.~\ref{phidot} to determine the sign of $\dot{\phi}$
at the midpoint of each solution.  Table \ref{Instanton Actions} indicates the Euclidean action integrals for the four instantons, with the limits
designating the direction of the path about the polar axis.

\begin{table}[!htbp]
\centering
\begin{tabular}{|c|c|}\hline
Instanton&Euclidean Action\\\hline
$1$&$iJ\int_{\pi/2}^{-\pi/2}\left(1-p_{+,-}(\phi)\right)d\phi$\\\hline
$2$&$iJ\int_{\pi/2}^{3\pi/2}\left(1-p_{-,-}(\phi)\right)d\phi$\\\hline
$3$&$iJ\int_{\pi/2}^{-\pi/2}\left(1-p^*_{-,-}(\phi)\right)d\phi$\\\hline
$4$&$iJ\int_{\pi/2}^{3\pi/2}\left(1-p^*_{+,-}(\phi)\right)d\phi$\\\hline
\end{tabular}
\caption{Instanton Actions: the upper bound $-\pi/2$ versus $3\pi/2$ is intended to convey whether the instanton runs through $0$ or $\pi$.}
\label{Instanton Actions}
\end{table}

By symmetry, only one of these integrals need actually be evaluated.  The contour in the complex-$\phi$ plane associated with one of these solutions
(Instanton
1) is shown in Fig.~\ref{ParametricG3}.  That the contour begins and ends at a branch point poses no complications since the integrand vanishes
smoothly at those points.  With the branch cuts shown, the contour can be continuously deformed onto (or within $\varepsilon$ of) the real-$\phi$
axis.
Along this axis, the real part of $p_{\pm,-}$ (the part we are interested in) is only non-zero between the inner two branch points.
The imaginary part of the Euclidean action (for Instanton 1) can then be written as
\begin{eqnarray}
S_I&=&-J\left(\pi + I\right)\nonumber \\
I&=&\frac{1}{J}\int_{b}^{-b}\text{Re}\left[p_{+,-}(\phi)\right]d\phi
\label{simag},
\end{eqnarray}
where $\pm b$ are the positions of the inner branch points.  Proceeding similarly for the other instantons, the total tunnel splitting can be
calculated using Eq.~(\ref{instanton splitting}):
\begin{eqnarray}
\Delta(\lambda)
&\propto& 
e^{-S_R}[e^{-iJ\pi}\left(e^{-iJI}+e^{iJI}\right)\nonumber \\
& &+e^{iJ\pi}\left(e^{-iJI}+e^{iJI}\right)] \nonumber \\
&=&4e^{-S_R}\cos\left(J\pi\right)\cos\left(JI\right), \label{final tunnel splitting}
\end{eqnarray}
where $S_R$ is the real part of the Euclidean action and the same for all instantons. The factor $\cos\left(J\pi\right)$ ensures that the tunnel
splitting is zero for half-integer spin, in accordance with Kramers' theorem \cite{149, 150}.  It arises from the interference between diametrically
opposite paths (Fig.~\ref{Effects of Lambda}) that remains constructive (destructive) for integer (half-integer) values of $J$ as $\lambda$ is varied.
The $\cos\left(JI\right)$ factor is of chief concern here.  Since the integral $I$ depends on anisotropy parameters $a$ and $\lambda$, variations in
these parameters can modulate the tunnel splitting, with complete suppression of tunneling when $JI=(2n+1)\pi/2$ for integer $n$.  For fixed $a$, as
$\lambda$ is increased from zero, the inner branch points move towards each other and the value of $I$ decreases,
resulting in repeated suppression of tunneling until $\lambda=\lambda_c\equiv 4\sqrt{a-2a^2}$ at which point $b=0$ and the value of $I$ goes to zero.

\begin{figure}[htbp]
\centering
\includegraphics[width=0.40\textwidth]{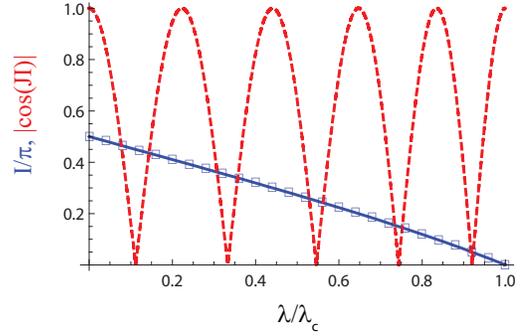}
\caption{(Color online) I/$\pi$ (blue, solid curve and points) and $|\cos\left(JI\right)|$ (red, dashed) as a function of $\lambda/\lambda_c$ for
$a=\frac{2-\sqrt{2}}{8}$ and $J=10$.  The points are the result of numerical evaluation of $I$ whereas the solid blue curve is obtained using
Eq.~(\ref{I}). The value of $a$ used is the largest one for which the problem is well defined for all values of $\lambda<\lambda_c$ (see Appendix).}
\label{IvsLambda}
\end{figure}

An approximate expression for $I$ can be obtained by treating the integrand as a function of two independent parameters ($a$ and $\lambda/\lambda_c$)
and expanding in a power series in $a$.  The resulting integrals can be done by elementary means.  We obtain
\begin{equation}
I=\frac{\pi}{2} \left(1 - \frac{\lambda}{\lambda_c}\right) g\left(a,\lambda \right),
\label{I}
\end{equation}
where $g\left(a,\lambda \right)$, up to order $a^2$, is
\begin{eqnarray}
g\left(a,\lambda \right)&=&1 + \frac{\lambda}{\lambda_c}\left(1 + \frac{\lambda}{\lambda_c}\right) \nonumber \\
& & \times\left(\frac{3 a}{2} -
\frac{a^2}{16} \left(9 - 133 \left(\frac{\lambda}{\lambda_c}\right)^2\right)\right).
\label{g}
\end{eqnarray}
Fig.~\ref{IvsLambda} compares the results of Eqs.~(\ref{I}) and (\ref{g}) (solid blue curve) to exact numerical evaluation of $I$ (points) for
$a=\frac{2-\sqrt{2}}{8}\approx0.073$, the largest value of $a$ for which $I$ can be meaningfully evaluated for all $\lambda<\lambda_c$ (see Appendix).  Even at this
relatively large value for $a$, Eq.~(\ref{I}) is clearly an excellent approximate expression for $I$, although for such large values of $a$ and
$\lambda$
the instanton approximation is questionable since the barrier has nearly disappeared.  The $\lambda$-dependence of $I$ is clearly dominated by the
$\left(1 - \frac{\lambda}{\lambda_c}\right)$ factor in Eq.~(\ref{I}), responsible for the nearly linear behavior seen in the figure.  To illustrate the
geometric interference effect, the figure also shows the behavior of the magnitude of the $\cos\left(JI\right)$ factor in $\Delta$ as a function of
$\lambda$ (red dashed curve) for $J=10$.  We note that when $J$ is an odd integer, the zeros in $\Delta$ start at $\lambda=0$ (because $I=\pi/2$ at
this point), as expected by general symmetry considerations \cite{150}.  Since the argument of $\cos\left(JI\right)$ ranges from zero to $J\pi/2$, the number of zeros
is $\lfloor(J+1)/2\rfloor$.

We next investigated the effect of a longitudinal field on the geometric-phase interference.  Such a field introduces a term $-h_z S_z$ into the
Hamiltonian, breaking the symmetry between up and down directions of spin (i.e. raising the energy of one minimum with respect to the other) but preserving  other rotational and reflection symmetries.  In particular,
the four instanton paths maintain the same amplitude and will continue to interfere.

\begin{figure}[htbp]
\centering
\includegraphics[width=0.40\textwidth]{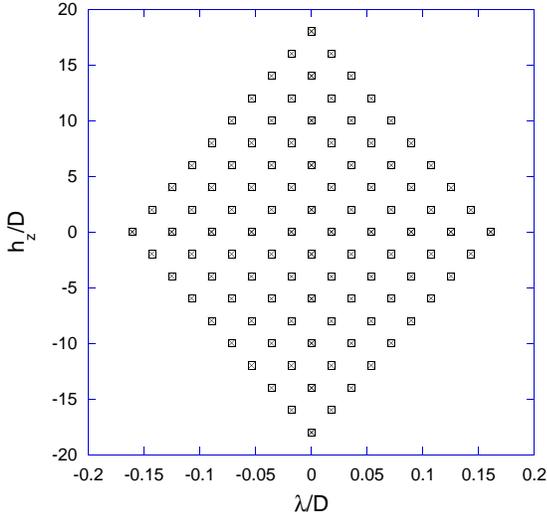}
\caption{Diabolical points in $\lambda-h_z$ parameter space with $k=2\times 10^{-5}D$ and $J=10$ (i.e. Mn$_{12}$) calculated using Eq.~(\ref{d points})
(squares)
and by exact diagonalization of the Hamiltonian ($\times$'s).  Points at negative values of $h_z$ and $\lambda$ were obtained by reflection in the
horizontal and/or vertical axis.}
\label{diabolical}
\end{figure}

Similar to what was found for the biaxial spin, we determined that a spin described by Eq.~\ref{Ham} has an array of diabolical points.  At certain values of longitudinal field, varying
$\lambda$ causes the tunnel splitting to oscillate, with perfect degeneracies occurring at a discrete set of points in $\lambda-h_z$ space.
The effect can be most readily  addressed by invoking the argument of Bruno \cite{607} that tunneling between  $|J\rangle$ and  $|-J+n\rangle$ can be
mapped approximately onto the problem of tunneling between the ground states $|\tilde{J}\rangle$ and  $|-\tilde{J}\rangle$ of an effective spin
$\tilde{J}=J-n/2$ in zero magnetic field.  Making the substitution $J \rightarrow \tilde{J}$ in Eqs.~(\ref{final tunnel splitting}) and (\ref{I}) and
keeping in mind that both $a$ and $\lambda_c$ are functions of $J$, we can generate the full set of diabolical points.  If $a$ is small, we can
approximate $g\left(a,\lambda \right)\approx 1$ and get the following simple expressions for the diabolical points.

\begin{eqnarray}
h_z&=&nD, \nonumber\\
\lambda&=&4\sqrt{kD-2(k\tilde{J})^2}\left[\tilde{J}-2l-1\right],
\label{d points}
\end{eqnarray}
where $l=0,1,\ldots, \lfloor(\tilde{J}-1)/2\rfloor$ and $n$ is an even (odd) integer when $J$ is a (half-)integer.  For other integer values of $n$,
Eq.~(\ref{final tunnel splitting}) yields exactly zero tunnel splitting for all values of $\lambda$, as expected in the absence of a transverse field.
In these expressions, we have restored the anisotropy parameter $D$. The first expression is equivalent to the condition for resonant tunneling \cite{33} in which the ground state associated with the up direction is resonant with the $n$th excited state in the down direction. Fig.~\ref{diabolical} shows a comparison between the results of Eq.~(\ref{d
points})
and exact numerical diagonalization of the Hamiltonian using $J=10$ and $k/D=2\times 10^{-5}$, a value close to that of Mn$_{12}$.  The agreement is obviously
very good.

We note that the pattern of diabolical points in Fig.~\ref{diabolical} resembles the pattern found in the biaxial system \cite{601, 248, 609, 592,kg2,607, 648}, where the diabolical points lie on a perfect lattice in $h_x-h_z$ space.  In contrast, the diabolical points in Fig.~\ref{diabolical} lie in the $\lambda-h_z$ plane and only form a perfect lattice in the limit of $k\rightarrow0$.

Finally, we consider the feasibility of observing the proposed effect.  For Mn$_{12}$Ac, $a\approx 3 \times 10^{-3}$, which gives $\lambda_c\approx 0.2 D = 0.11$ K.  Density functional calculations that include molecules of solvation in a trans configuration estimate that a 3\% strain along the axis defined by the solvent molecules can produce changes in $\lambda$ as large as 19 mK \cite{kyungwha}.  To the best of our knowledge, the elastic modulus of Mn$_{12}$Ac has not been measured, but it is unlikely that it could be much larger than 100 kbar.  One might then need uniaxial pressure as high as 17 kbar to reach $\lambda_c$, although the interference effect would become manifest at substantially lower values of $\lambda$.  Several experimental studies have investigated the effect of hydrostatic pressure on Mn$_{12}$ \cite{282, 488, 651, 650} and have found significant changes in $D$ with pressures as high as 12 kbar.  None of these investigations involved uniaxial pressure, which would be needed to observe the interference effect predicted herein.  We note that the ground-state tunnel splitting for Mn$_{12}$Ac is extremely small.  However, the splitting increases exponentially for progressively higher levels, suggesting that a pressure-induced effect would be observable in thermally excited states, i.e. by studying the magnetic relaxation in the thermally assisted tunneling regime \cite{33} as a function of pressure.  Discussion of the geometric-phase effect in excited levels will be presented elsewhere.

In summary, we predict a geometric-phase interference effect
involving four instanton paths that can be modulated by the strength of the second-order
transverse anisotropy, leading to a nearly periodic suppression of tunneling in the absence of any external field.  In principle, this effect can be
experimentally realized through the application of uniaxial stress along one of the hard-axis directions of a four-fold symmetric single-molecule
magnet, such as Mn$_{12}$.

\section{Appendix}
Here we discuss how the energy landscape for a spin governed by Eq.~\ref{Ecl} depends on the anisotropy parameters $a$ and $\lambda$.  (As in the main text, we measure these anisotropy parameters in units of $D$ and define $a\equiv kJ^2$.)  Using standard techniques from multivariable calculus we determined the minima, maxima and saddle points of the landscape.

\begin{figure}[htbp]
\centering
\includegraphics[width=0.40\textwidth]{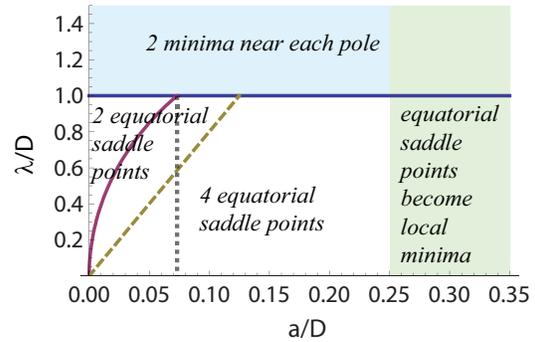}
\caption{Parameter-space plot indicating the essential features of the spin's energy landscape.  The unshaded region represents the locus of parameters to which the analysis in this paper applies.  In the shaded region on the right, the saddle points on the equator have become local minima.  In the shaded region above $\lambda=1$, the minima at the poles have bifurcated.  The dashed line indicates where equatorial saddle points merge pairwise.  The solid curve shows the behavior of $\lambda_c$ and is included for reference.}
\label{pspace}
\end{figure}

For small values of $a$ and $\lambda$ (unshaded region in Fig.~\ref{pspace}), the landscape resembles that shown in Fig.~\ref{Effects of Lambda} with global minima at the poles ($\pm z$ directions) and saddle points along the equator ($x-y$ plane).  As $a$ increases, the saddle points become deeper and for $a>1/4$ the saddle points transform into local minima on the equator (right shaded region in Fig.~\ref{pspace}).  This limit is far from relevant for Mn$_{12}$, where $a\approx 3 \times 10^{-3}$.

 Increasing $\lambda$ causes pairs of saddle points on the equator to move towards each other (see Fig.~\ref{Effects of Lambda}).  They merge together when $\lambda=8a$ (dashed line in Fig.~\ref{pspace}).  Remarkably, even when there are only two saddle points in the potential landscape, there remain four distinct instanton paths.  The solid angle between pairs of paths vanishes when $\lambda=\lambda_c$ (solid curve), which does not correspond to any notable change in the energy landscape.  The vertical dashed line represents the values in this parameter space used for the results presented in Fig.~\ref{IvsLambda}, i.e. $a$ is such that $\lambda_c=1$.

Finally, when $\lambda>1$ each energy minimum at a pole bifurcates into two minima, one tilted towards the +y direction and the other towards the -y direction.  This substantially changes the nature of the problem.  The analysis described in this paper applies only to the unshaded region in  Fig.~\ref{pspace} and would need modification to handle the energy landscapes corresponding to the shaded regions.

We thank W. Loinaz, K. Jagannathan, R. Benedetto, D. Velleman, R. Behrend and E. H. da Silva Neto for useful discussions.
Support for this work was provided by the U.S. National Science Foundation under grant no.~DMR-0449516.

\end{document}